\documentclass[preprint,showpacs,preprintnumbers,amsmath,amssymb]{revtex4}
\usepackage{graphicx}
\usepackage{dcolumn}
\usepackage{bm}

\newcommand{\be}{\begin{equation}}
\newcommand{\en}{\end{equation}}
\newcommand{\bea}{\begin{eqnarray}}
\newcommand{\ena}{\end{eqnarray}}

\begin{document}


\title{Tachyon-Chaplygin inflation on the brane  }

\author{Ram\'on Herrera}
\email{ramon.herrera@ucv.cl} \affiliation{ Instituto de
F\'{\i}sica, Pontificia Universidad Cat\'{o}lica de
Valpara\'{\i}so, Casilla 4059, Valpara\'{\i}so, Chile.}

\date{\today}

\begin{abstract}
 Tachyon-Brane inflationary universe model in the context of a  Chaplygin
 gas equation of state is studied. General
 conditions  for this model to be realizable are
 discussed. In the high-energy limit and by using an exponential potential we
 describe in great details the characteristic of this model.
 Recent observational data from the Wilkinson Microwave Anisotropy
 Probe experiment are employed to restrict the parameters of the
 model.
\end{abstract}

\maketitle

\section{Introduction}

Inflationary universe models have solved many problems of the
Standard Hot Big Bang scenario, for example, the flatness, the
horizon, and the monopole problems, among others
\cite{guth,infla}. In addition, its has provided a causal
interpretation of the origin of the observed anisotropy of the
cosmic microwave background (CMB) radiation, and also the
distribution of large scale structures.\cite{astro}.

In concern to higher dimensional theories, implications of
string/M-theory to Friedmann-Robertson-Walker (FRW) cosmological
models have recently attracted  great deal of attention, in
particular, those related to brane-antibrane configurations such
as space-like branes\cite{sen1}. The realization  that we may live
on a so-called brane embedded in a higher-dimensional Universe has
significant implications to cosmology \cite{1}. In this scenario
the standard model  particles are confined on the  brane, while
gravitations propagate in the bulk spacetimes. Since, the effect
of the extra dimension induces additional terms in the Friedmann
equation is modified at very high energies \cite{22,3}, acquiring
a  quadratic term in the energy density. Such a term generally
makes it easier to obtain inflation in the early Universe
\cite{4,5}. For a review, see, e.g., Ref.\cite{M}.

In recent times a great amount of work has been invested in
studying the inflationary model with a tachyon field. The tachyon
field associated with unstable D-branes might be responsible for
cosmological inflation in the early evolution of the universe, due
to tachyon condensation near the top of the effective scalar
potential \cite{sen2}, which could
 also add some new form of cosmological dark matter at late times
\cite{Sami_taq}. Cosmological implications of this rolling tachyon
were first studied by Gibbons \cite{gibbons} and in this context
it is quite natural to consider scenarios in which inflation is
drive by the rolling tachyon.

On the other hand, the generalized Chaplygin gas    has been
proposed as an  alternative model for  describing the accelerating
of the universe. The generalized Chaplygin gas is described by an
exotic equation of state of the form $ p_{ch} = -
A\,\rho_{ch}^{-\beta}$, where $\rho_{ch}$ and $p_{ch}$ are the
energy density and pressure of the generalized Chaplygin gas,
respectively \cite{Bento}. $\beta$ is a constant that lies in  the
range $0 <\beta\leq 1$, and $A$ is a positive constant. The
original  Chaplygin gas corresponds to the case $\beta = 1$
\cite{2}. Inserting this equation of state into the relativistic
energy conservation equation leads to an energy density given by
\cite{Bento}
\begin{equation}
 \rho_{ch}=\left[A+\frac{B}{a^{3(1+\beta)}}\right]^{\frac{1}{1+\beta}},
 \label{2}
\end{equation}
 where $a$ is the
scale factor  and $B$ is a positive integration constant.

The Chaplygin gas emerges as a effective fluid of a generalized
d-brane in a (d+1, 1) space time, where the   action can be
written as a generalized Born-Infeld action \cite{Bento}. These
models have been extensively studied in the literature
\cite{other,CMB,SIa}.

In the  model of  Chaplygin inspired inflation usually the scalar
field, which drives inflation, is the standard inflaton field,
where the energy density given by Eq.(\ref{2}), can be extrapolate
for  obtaining a successful inflation period with a Chaplygin gas
model\cite{Ic}. Recently, tachyon-Chaplygin inflationary universe
model was considered in \cite{yo}, and the dynamics of the early
universe and the initial conditions for inflation in a model with
radiation and a Chaplygin gas was studied  in
Ref.\cite{Monerat:2007ud} (see also \cite{yoyo} ).

A natural extension of Ref.\cite{yo} is to consider the tachyon
field as a degree of freedom on visible three dimensional brane.
This work has been extended to include higher order corrections in
slow-roll parameters  and the formula has been widely used to
confront this model with the observations. Moreover, we find
constraints on the parameter $A$ and the five-dimensional Planck
mass or equivalently the brane tension.


The outline of the paper is a follows. The next section presents a
short review of the modified Friedmann equation by using a
Chaplygin gas, and  we present the tachyon-brane-Chaplygin
inflationary model. Section \ref{sectpert} deals with the
calculations of cosmological perturbations in general term.  In
Section \ref{exemple} we use an exponential  potential in the
high-energy limit for obtaining explicit expression for the model.
Finally, Sect.\ref{conclu} summarizes our findings.

\section{The modified Friedmann equation and the tachyon-brane-Chaplygin Inflationary phase. }

 We consider the five-dimensional brane scenario, in which the
Friedmann equation is modified from its usual form, in the
following way\cite{2,3}
\begin{equation}
H^2=\kappa\,\rho_\phi\left[1+\frac{\rho_\phi}{2\lambda}\right]+\frac{\Lambda_4}{3}+\frac{\xi}{a^4},
\end{equation}
where  $H=\dot{a}/a$ denotes the Hubble parameter, $\rho_\phi$
represents the matter confined to the brane, $\kappa=8\pi
G/3=8\pi/3m_p^2$ ($m_p$ is the four-dimensional Planck mass),
$\Lambda_4$ is the four-dimensional cosmological constant and  the
final term represents the influence of bulk gravitons on the
brane, where $\xi$ is an integration constant (this term appears
as a form of dark radiation). The brane tension $\lambda$ relates
the four and five-dimensional Planck masses via
$m_p=\sqrt{3M_5^6/(4\pi\lambda)}$, and is constrained by the
requirement of successful nucleosynthesis as $\lambda >$
(1MeV)$^4$ \cite{Cline}. We assume that the four-dimensional
cosmological constant is set to zero, and once inflation begins
the final term will rapidly become unimportant, leaving us
with\cite{M}. Hence, the modified Friedmann equation reads

\begin{equation}
H^2=\kappa\left[A+\rho_\phi^{(1+\beta)}\right]^{\frac{1}{1+\beta}}\,
\left[1+\frac{\left[A+\rho_\phi^{(1+\beta)}\right]^{\frac{1}{1+\beta}}}{2\lambda}\right].
\label{HC}
\end{equation}
Here, $\rho_\phi$  becomes
$\rho_\phi=V(\phi)/\sqrt{1-\dot{\phi}^2}$,
  and $V(\phi)=V$ is
the tachyonic potential. Note that, in the low energy regime
$[A+\rho_\phi^{(1+\beta)}]^{1/(1+\beta)}\ll\lambda$, the
tachyon-Chaplygin inflationary model is recovered \cite{yo}, and
in a very hight-energy regime, the contribution from the matter in
Eq.(\ref{HC})  becomes proportional to
$[A+\rho_\phi^{(1+\beta)}]^{2/(1+\beta)}$ in the effective  energy
density.

 We assume that the tachyon field is confined
to the brane, so that its field equation has the  form
 \be \frac{\ddot{\phi}}{1-\dot{\phi}^2}\,+3H \;
\dot{\phi}+\frac{V'}{V}=0, \label{key_01}
 \en
where dots mean derivatives with respect to the cosmological time
and
 $V'=\partial V(\phi)/\partial\phi$. For convenience we will use
 units in which $c=\hbar=1$.

  The modification of the Eq.(\ref{HC}) is realized from an
extrapolation of Eq.(\ref{2}), where  the density matter
$\rho_m\sim a^{-3}$ in introduced in such  a way that we may write
$\rho_{ch}=\left[A+\rho_m^{(1+\beta)}\right]^{\frac{1}{1+\beta}}\longrightarrow
 \left[A+\rho_\phi^{(1+\beta)}\right]^{\frac{1}{1+\beta}},
$
and  thus, we identifying $\rho_m$ with the contributions of the
scalar tachyon field which gives Eq.(\ref{HC}). The generalized
Chaplygin gas model may be viewed as a modification of gravity, as
described in Ref.\cite{Ber},  for chaotic inflation, in
Ref.\cite{Ic}, and for tachyon-Chapligin inflationary universe
model in the low-energy limit, in Ref.\cite{yo}. Different
modifications of gravity have been proposed in the last few years,
and there has been a lot of interest in the construction of early
universe scenarios in higher-dimensional models motivated by
string/M-theory \cite{Ran}. It is  well-known that these
modifications can lead to important changes in the early universe.
In the following  we will take $\beta=1$ for simplicity, which
means the usual Chaplygin gas.

During the inflationary epoch the energy density associated to the
 tachyon field is of the order of the potential, i.e.
$\rho_\phi\sim V$. Assuming the set of slow-roll conditions, i.e.
$\dot{\phi}^2 \ll 1$ and $\ddot{\phi}\ll 3H\dot{\phi}$
\cite{gibbons,Fairbairn:2002yp}, the Friedmann equation (\ref{HC})
reduces  to
\begin{eqnarray}
H^2\approx
\kappa\,\sqrt{A+V^2}\left[1+\frac{\sqrt{A+V^2}}{2\lambda}\right],\label{inf2}
\end{eqnarray}
and  Eq. (\ref{key_01}) becomes
\begin{equation}
3H \dot{\phi}\approx-\frac{V'}{V}. \label{inf3}
\end{equation}

Introducing the dimensionless slow-roll parameters
\cite{Hwang:2002fp}, we write
\begin{equation}
\varepsilon=-\frac{\dot{H}}{H^2}\simeq\frac{m_p^2}{16\pi}\,\left[\frac{V'^2}{(A+V^2)^{3/2}}
\,\frac{\left(1+\frac{(A+V^2)^{1/2}}{\lambda}\right)}{\left(1+\frac{(A+V^2)^{1/2}}{2\lambda}\right)^2}\right],\label{ep}
\end{equation}
\begin{equation}
\eta=-\frac{\ddot{\phi}}{H\,
\dot{\phi}}\simeq\,\frac{m_p^2}{8\pi}\,\left(\frac{V''}{V\;(A+V^2)^{1/2}}\right)\,\left[1+\frac{(A+V^2)^{1/2}}{2\lambda}
\right]^{-1}\label{eta},
\end{equation}
and
\begin{equation}
\gamma=-\frac{V'\,\dot{\phi}}{2\,H\,V}\simeq\frac{m_p^2}{16\,\pi}\,\left(\frac{V'\,^2}{V^2\,(A+V^2)^{1/2}}\right)\;
\left[1+\frac{(A+V^2)^{1/2}}{2\,\lambda}\right]^{-1}.
\end{equation}

Note that  in the low-energy limit,
$\sqrt{A+\rho_\phi^2}\ll\lambda$, the slow-parameters are
recovered \cite{yo}.

The condition under  which inflation  takes place can be
summarized with the parameter $\varepsilon$ satisfying  the
inequality $\varepsilon<1$, which  is analogue to the requirement
that  $\ddot{a}> 0$. This condition could be written in terms of
the  tachyon potential and its  derivative $V'$, which becomes
\begin{equation}
V'^2\,\left[1+\frac{(A+V^2)^{1/2}}{\lambda}\right]<\frac{16\pi}{m_p^2}\,(A+V^2)^{3/2}\,
\left[1+\frac{(A+V^2)^{1/2}}{2\lambda}\right]^2.\label{cond}
\end{equation}

Inflation ends when the universe heats up at a time when
$\varepsilon\simeq 1$, which implies
\begin{equation}
V_f'^2\,\left[1+\frac{(A+V_f^2)^{1/2}}{\lambda}\right]\simeq\frac{16\pi}{m_p^2}\,(A+V_f^2)^{3/2}\,
\left[1+\frac{(A+V_f^2)^{1/2}}{2\lambda}\right]^2.\label{al}
\end{equation}
However, in the high-energy limit
$[A+\rho_\phi^2]^{1/2}\approx[A+V^2]^{1/2}\gg\lambda$
Eq.(\ref{al}) becomes
$$
V_f'^2\simeq\frac{4\pi}{m_p^2}\,\frac{(A+V_f^2)^2}{\lambda}.
$$

The number of e-folds at the end of inflation is given by
\begin{equation}
N=-\frac{8\pi}{m_p^2}\,\int_{\phi_{*}}^{\phi_f}\frac{V\;\sqrt{A+V^2}}{V'}
\,\left[1+\frac{\sqrt{A+V^2}}{2\lambda}\right]\,d\phi,\label{N}
\end{equation}
or equivalently
\begin{equation}
N=-\frac{8\pi}{m_p^2}\,\int_{V_{*}}^{V_f}\frac{V\;\sqrt{A+V^2}}{V'^2}\left[1+\frac{\sqrt{A+V^2}}{2\lambda}\right]
\,d\,V.\label{NV}
\end{equation}
Note that in the high-energy limit Eq.(\ref{NV}) becomes
$N\simeq-(4\pi/m_p^2\lambda)\int_{V_*}^{V_f}\,[V\;(A+V^2)/V'^2]dV$.

  In the following, the subscripts  $*$ and $f$ are
used to denote  the epoch when the cosmological scales exit the
horizon and the end of  inflation, respectively.

\section{Perturbations\label{sectpert}}

In this section we will study the scalar and tensor perturbations
for our model. It was shown in Ref. \cite{PB} that the
conservation of the curvature perturbation, $\cal{R}$, holds for
adiabatic perturbations irrespective of the form of gravitational
equations by considering the local conservation of the
energy-momentum tensor. However, we note here that even though the
effect of bulk to the cosmological perturbations can not be
trivially negligible, it can be shown that the main correction of
the spectrum in the brane-world inflation is just the modification
of the slow-roll parameters\cite{YOK} (see also \cite{YOK1}).
 For a tachyon field the power spectrum of
the curvature perturbations  is given
${\cal{P}_R}\simeq\left(\frac{H^2}{2\pi\dot{\phi}}\right)^2\,\frac{1}{Z_s}$,
where $Z_s=V\,(1-\dot{\phi}^2)^{-3/2}$ \cite{Hwang:2002fp}. Under
the slow-roll approximation, the power spectrum of curvature
perturbations is estimated to be \cite{Hwang:2002fp}

\begin{equation}
{\cal{P}_R}\simeq\left(\frac{H^2}{2\pi\dot{\phi}}\right)^2\,
\frac{1}{V}\simeq\,\frac{128\pi}{3m_p^6}\,\left(\frac{V\,(A+V^2)^{3/2}}{V'^2}\right)\,
\left[1+\frac{(A+V^2)^{1/2}}{2\lambda}\right]^3 .\label{dp}
\end{equation}
Note that in the low-energy limit  the amplitude of scalar
perturbation given by Eq.(\ref{dp}) coincides with Ref.\cite{yo}.

The scalar spectral index $n_s$ is given by $ n_s -1 =\frac{d
\ln\,{\cal{P}_R}}{d \ln k}$,  where the interval in wave number is
related to the number of e-folds by the relation $d \ln k(\phi)=-d
N(\phi)$. From Eq.(\ref{dp}), we get,  $n_s  \approx\,
1\,-2(2\varepsilon+\gamma-\eta)$,
 or equivalently
$$
n_s  \approx\,
1\,-\frac{m_p^2}{4\pi}\,(A+V^2)^{-1/2}\,\left[1+\frac{(A+V^2)^{1/2}}{2\lambda}\right]^{-1}\,\times
$$
\begin{equation}
\left(\frac{V'\,^2}{(A+V^2)}\,\frac{\left[1+\frac{(A+V^2)^{1/2}}{\lambda}\right]}
{\left[1+\frac{(A+V^2)^{1/2}}{2\lambda}\right]}+\frac{V'\,^2}{2\,V^2}-\frac{V''}{V}\right).\label{nsa}
\end{equation}

One of the interesting features of the five-year data set from
Wilkinson Microwave Anisotropy Probe (WMAP) is that it hints at a
significant running in the scalar spectral index $dn_s/d\ln
k=\alpha_s$ \cite{astro}. From Eq.(\ref{nsa}) we get  that the
running of the scalar spectral index becomes

\begin{equation}
\alpha_s=\left(\frac{4\,(A+V^2)}{V\;V'}\right)\,\left[\frac{\left(1+\frac{(A+V^2)^{1/2}}{2\lambda}\right)}
{\left(1+\frac{(A+V^2)^{1/2}}{\lambda}\right)}\right]\;[2\,\varepsilon_{,\,\phi}+\gamma_{,\,\phi}-\eta_{,\,\phi}]
\;\varepsilon.\label{dnsdk}
\end{equation}
In models with only scalar fluctuations the marginalized value for
the derivative of the spectral index is approximately $-0.03$ from
WMAP-five year data only \cite{5}.

On the other hand, the generation of tensor perturbations during
inflation would produce  gravitational waves and this
perturbations in cosmology are more involved since gravitons
propagate in the bulk. The amplitude of tensor perturbations was
evaluated in Refs.\cite{t} and \cite{Sami:2004rk}
\begin{equation}
{\cal{P}}_g=24\kappa\,\left(\frac{H}{2\pi}\right)^2\;F^2(x)
\simeq\frac{6}{\pi^2}\,\kappa^2\,(A+V^2)^{1/2}\,\left[1+\frac{(A+V^2)^{1/2}}{2\lambda}\right]\,F^2(x),\label{ag}
\end{equation}
where $x=Hm_p\sqrt{3/(4\pi\lambda)}$ and
$$
F(x)=\left[\sqrt{1+x^2}-x^2\sinh^{-1}(1/x)\right]^{-1/2}.
$$
Here the function $F(x)$ appeared from the normalization of a
zero-mode. The spectral index $n_g$ is given by $
n_g=\frac{d{\cal{P}}_g}{d\,\ln
k}=-\frac{2x_{,\,\phi}}{N_{,\,\phi}\,x}\frac{F^2}{\sqrt{1+x^2}}$.

From expressions (\ref{dp}) and (\ref{ag}) we write  the
tensor-scalar ratio as
\begin{equation}
r(k)=\left.\left(\frac{{\cal{P}}_g}{P_{\cal
R}}\right)\right|_{k=k_*} \simeq
\left.\left(\frac{8}{3\,\kappa}\,\frac{V'^2\;F^2(V)}{V\;(A+V^2)\;[1+(A+V^2)^{1/2}/2\lambda]^2}\right)\right|_{\,k=k_*}.
\label{Rk}\end{equation} Here, $k_*$  is referred to $k=Ha$, the
value when the universe scale  crosses the Hubble horizon  during
inflation.

Combining  WMAP five-year data\cite{astro} with the Sloan Digital
Sky Survey  (SDSS) large scale structure surveys \cite{Teg}, it is
found an upper bound for $r$ given by $r(k_*\simeq$ 0.002
Mpc$^{-1}$)$ <0.28\, (95\% CL)$, where $k_*\simeq$0.002 Mpc$^{-1}$
corresponds to $l=\tau_0 k\simeq 30$,  with the distance to the
decoupling surface $\tau_0$= 14,400 Mpc. The SDSS  measures galaxy
distributions at red-shifts $a\sim 0.1$ and probes $k$ in the
range 0.016 $h$ Mpc$^{-1}$$<k<$0.011 $h$ Mpc$^{-1}$. The recent
WMAP five-year results give the values for the scalar curvature
spectrum $P_{\cal R}(k_*)\simeq 2.4\times\,10^{-9}$ and the
scalar-tensor ratio $r(k_*)=0.055$. We will make use of these
values  to set constrains on the parameters appearing in  our
model.

\section{Exponential potential in the high-energy limit. \label{exemple}}
Let us consider a  tachyonic effective potential $V(\phi)$,  with
the properties satisfying   $V(\phi)\longrightarrow$ 0 as
$\phi\longrightarrow \infty$. The exact form of the potential is
$V(\phi)=\left ( 1+\alpha \phi \right )\exp(-\alpha \phi)$, which
in the case when $\alpha \rightarrow 0$, we may use the asymptotic
exponential expression. This form for the potential is derived
from string theory calculations\cite{ku,sen2}. Therefore, we
simple use
\begin{equation}
V(\phi)=V_0 e^{-\alpha\phi},\label{pot}
\end{equation}
where $\alpha$ and $V_0$ are free parameters. In the following we
will restrict ourselves to the case in which  $\alpha> 0$. Note
that $\alpha$ represents the tachyon mass
\cite{Fairbairn:2002yp,delaMacorra:2006tm}. In Ref.\cite{Sami_taq}
is given an estimation of these parameters  in the low-energy
limit and  $A\rightarrow 0$. Here, it was found  $V_0\sim
10^{-10}m_p^4$ and $\alpha\sim 10^{-6} m_p$. In the following, we
develop models in the high-energy limit, i.e.
$\sqrt{A+V^2}\gg\lambda$.

From Eq.(\ref{NV}) the number of e-folds results in
\begin{equation}
N=\frac{4\pi}{\lambda\,\alpha^2\,m_p^2}\;[h(V_*)-h(V_f)],
\end{equation}
where
\begin{equation}
h(V)=\frac{V^2}{2}+A\;\ln V.
\end{equation}

On the other hand, we may establish that the end of  inflation is
governed by the condition  $\varepsilon=1$, from which we get that
the square of the tachyonic  potential becomes
\begin{equation}
V(\phi=\phi_f)^2=V_f^2=\frac{1}{8\,\pi}\;\left[\lambda\,\alpha^2\,m_p^2-8\,\pi\,A+
\sqrt{\lambda\,\alpha^2\,m_p^2(\lambda\,\alpha^2\,m_p^2-16\,\pi\,A)}\right],
\end{equation}
and
\begin{equation}
\dot{\phi}_f=\frac{\alpha\,m_p}{2\,V_f}\,\sqrt{\frac{\lambda}{3\,\pi}}.
\end{equation}

Note that in the limit $A\rightarrow 0$ we obtain $V_f=
\alpha\,m_p\,\sqrt{\lambda}/(2\sqrt{\pi})$ and
$\dot{\phi}_f=1/\sqrt{3}$, which coincides with that reported in
Ref.\cite{Sami_taq}.

From Eq.(\ref{dp}) we obtain that the scalar power spectrum is
given by
\begin{equation}
P_{\cal R}(k)\approx
\;\frac{16\pi}{3\,m_p^6}\,\frac{1}{\alpha^2\,\lambda^3}\left[\frac{(A+V^2)^{3}}{V}\right],\label{ppp}
\end{equation}
and from Eq.(\ref{Rk}) the tensor-scalar ratio becomes
\begin{equation}
r(k)\approx\;\frac{4\,m_p^2\,
\lambda^2\,\alpha^2}{\pi}\left[\frac{V}{(A+V^2)^{2}}\,F^2(V)\right].\label{rrrr}
\end{equation}

By using, that $V'=-\alpha\,V$, we obtain from Eq.(\ref{nsa})
\begin{equation}
n_s-1\simeq-\frac{m_p^2}{4\pi}\,\frac{\lambda\;\alpha^2}{(A+V^2)}\left[\frac{4\,V^2}{(A+V^2)}-1\right],\label{nV}
\end{equation}
and from  Eq.(\ref{dnsdk}) that
\begin{equation}
\alpha_s\simeq-\frac{\lambda^2\,\alpha^4\,m_p^4}{4\pi^2}\,\left[\frac{3\,V^2-5\,A}{(A+V^2)^4}\right]\,V^2.\label{as}
\end{equation}

The Eqs.(\ref{ppp}) and (\ref{nV})  has roots that can be solved
analytically for the parameters $\alpha$ and $A$, as a function of
$n_s$, $P_{\cal R}$, $V$ and $\lambda$. The real root solution for
$m^2$, and $A$ becomes
\begin{equation}
\alpha^2=2\,\pi\,\left[\frac{4^4\,V^5+6^2\,P_{\cal
R}\,(n_s-1)\,V^2\,\lambda^2\,m_p^4+\aleph\left(3\,P_{\cal
R}\,(1-n_s)\,\lambda^2\,m_p^4-8^2\,V^3\right)}{3\,P_{\cal
R}\,\lambda^3\,m_p^6} \right],\label{m}
\end{equation}
and
\begin{equation}
A=\frac{1}{2}\left(2\,V^2-\aleph\right)\label{A},
\end{equation}
 where
 $$
\aleph=\sqrt{16\,V^4+3(n_s-1)P_{\cal R}\,V\,\lambda^2\,m_p^4}\,.
 $$


From Eq.(\ref{A}) and  since $A>0$, the ratio $V^3/\lambda^2$
satisfies the inequality $V^3/\lambda^2<(1-ns)P_{\cal
R}\,m_p^4/4$. This inequality allows us to obtain an upper  limit
for the ratio $V^3(\phi)/\lambda^2$ evaluate when the cosmological
scales exit the horizon, i.e. $V_*^3/\lambda^2< 2.4\times
10^{-11}m_p^4$. Here, we have used the WMAP five year data where
$P_{\cal R}(k_*)\simeq 2.4\times 10^{-9}$ and $n_s(k_*)\simeq
0.96$.

One again, note that in the limit $A\rightarrow 0$,  the
constrains $\alpha\,\approx7\,\times10^{-3}M_5$ and
$V_*\approx4\times10^{-4}M_5^4$ are recovered \cite{Sami_taq}.
Here,  we used the relation $m_p=M_5^3\,\sqrt{3/(4\pi\lambda)}$.

 In Fig. 1 we
have plotted the adimensional quantity   $\lambda^2/m_p^8$ versus
the adimensional scalar tachyon potential evaluated when the
cosmological scales exist the horizon $V_*/m_p^4$. In doing this,
we using Eq.(\ref{as}) that has roos that can be solved for the
brane tension $\lambda$, as a function of $\alpha_s$, $m$, $A$ and
$V$. For a real root solution for $\lambda$, and from Eqs.
(\ref{m}) and (\ref{A}) we obtain a relation of the form
$\lambda=f(V_*)$ for a fixed  values of $\alpha_s$, $n_s$ and
$P_{\cal R}$. In this plot we using the WMAP five year data where
$P_{\cal R}(k_*)\simeq 2.4\times 10^{-9}$,  $n_s(k_*)\simeq 0.96$,
$\alpha_s(k_*)\simeq-0.03$. In Fig. 2 we have plotted the
tensor-scalar ratio given by Eq.(\ref{rrrr}) versus the
adimentional parameter $A/m_p^8$.  The WMAP five-year data favors
the tensor-scalar ratio $r\simeq 0.055$  and the from Fig. 2 we
obtain that $A$ parameter becomes $A\simeq
2.6\times10^{-25}m_p^8$. In Fig. 3 we have plotted the
tensor-scalar ratio given by Eq.(\ref{rrrr}) versus the
adimentional parameter $\alpha^2/m_p^2$. We note that for $r\simeq
0.055$ we obtain $\alpha^2\simeq 1.3\times10^{-12}m_p^2$.

For these values of the $A$ and $\alpha$ parameters we  get the
values $V_*\simeq1.3\times10^{-12}m_p^4$,
$V_f\simeq8.9\times10^{-14}m_p^4$ and   $\lambda\simeq 5.1\times
10^{-13}m_p^4\simeq4\times10^{-5}M_5^4$. Also,  the number of
e-folds, $N$, becomes of the order of $N\simeq 52.7$.  We should
note also that the $A$ parameter becomes smaller by two order of
magnitude and the $\alpha$ parameter becomes similar  when it are
compared with the case of tachyon-Chaplygin inflation in the
low-energy limit\cite{Ic}.

\begin{figure}[th]
\includegraphics[width=3.0in,angle=0,clip=true]{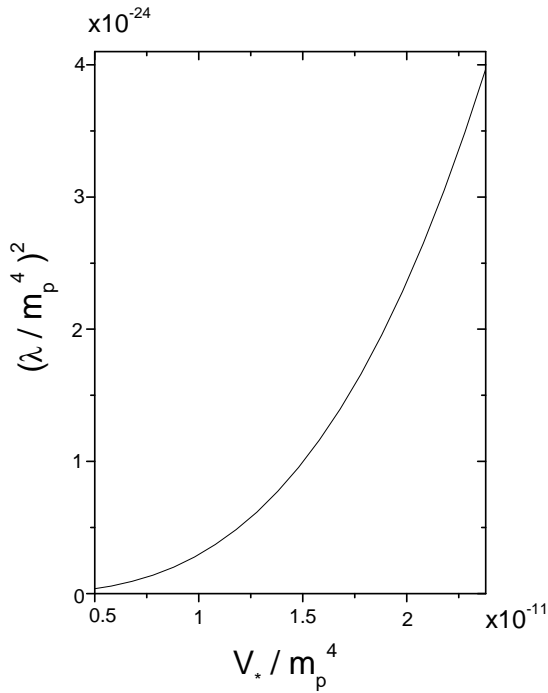}
\caption{The plot shows the adimentional square of the brane
tension $(\lambda/m_p^4)^2$ versus the adimentional scalar
potential
 $V_*/m_p^4$. Here, we have used the WMAP five-year data where
$P_{\cal R}(k_*)\simeq 2.4\times 10^{-9}$, $n_s(k_*)\simeq 0.96$
and $\alpha_s(k_*)\simeq-0.03$.
 \label{rons}}
\end{figure}

\begin{figure}[th]
\includegraphics[width=3.0in,angle=0,clip=true]{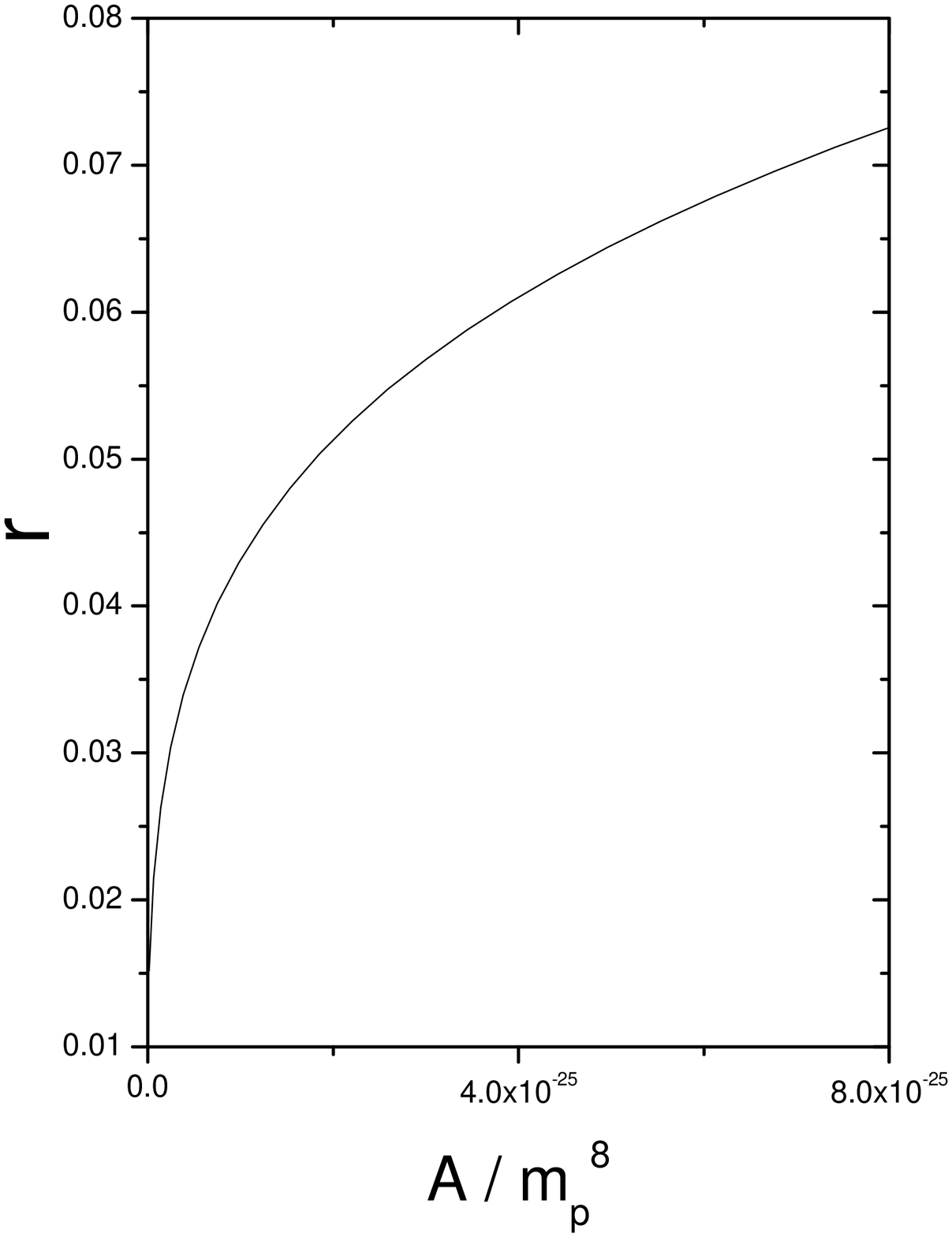}
\caption{The plot shows the tensor-scalar ratio $r$ versus the
adimentional parameter
 $A/m_p^8$. Here, we have used  the WMAP five-year data where
$P_{\cal R}(k_*)\simeq 2.4\times 10^{-9}$, $n_s(k_*)\simeq 0.96$
and $\alpha_s(k_*)\simeq-0.03$.
 \label{rons}}
\end{figure}

\begin{figure}[th]
\includegraphics[width=3.0in,angle=0,clip=true]{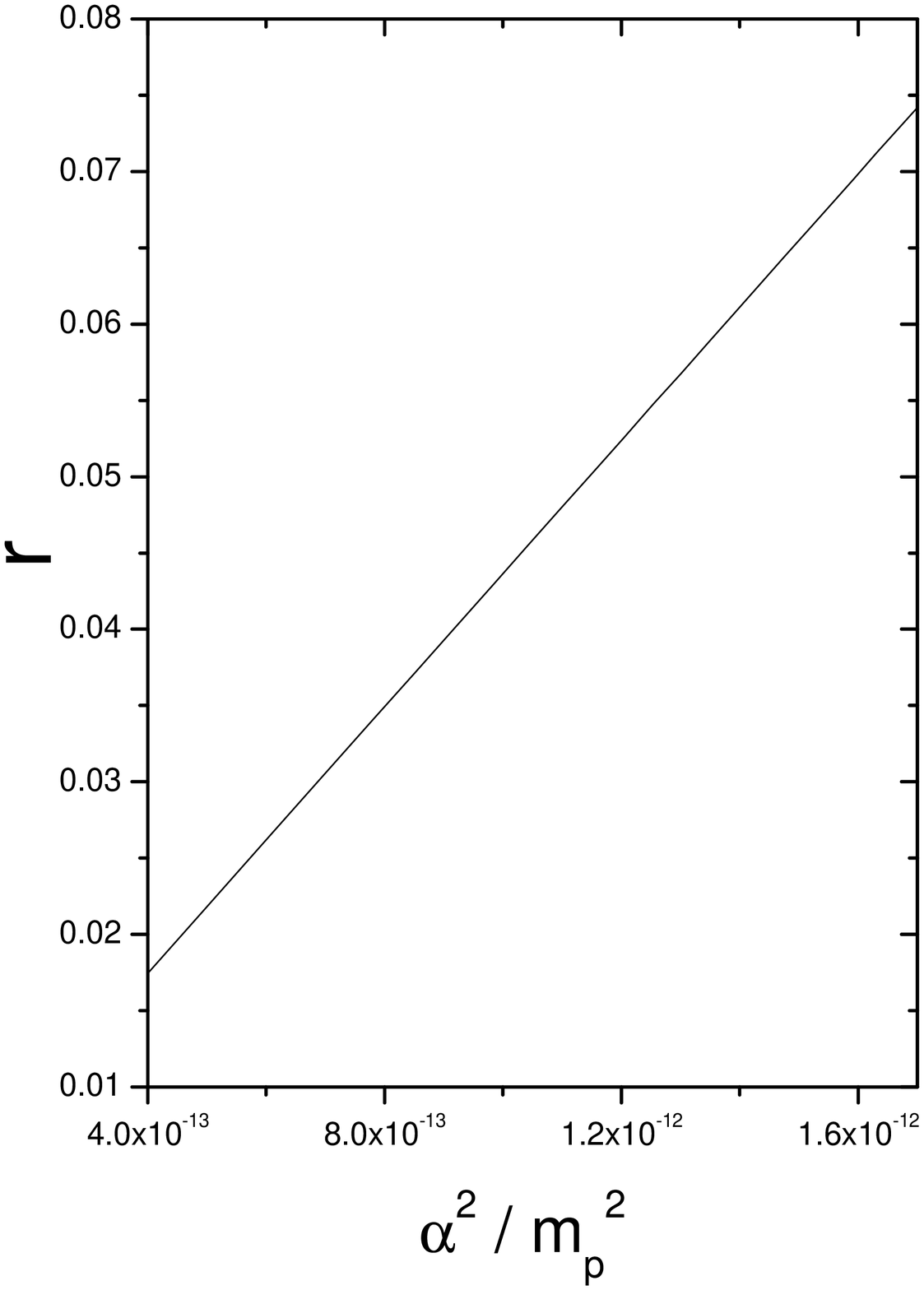}
\caption{The plot shows the tensor-scalar ratio $r$ versus the
adimentional parameter
 $\alpha^2/m_p^2$. Here, we have used  the WMAP five-year data where
$P_{\cal R}(k_*)\simeq 2.4\times 10^{-9}$, $n_s(k_*)\simeq 0.96$
and $\alpha_s(k_*)\simeq-0.03$.
 \label{rons}}
\end{figure}


\section{Conclusions \label{conclu}}

In this work, we have studied the tachyon-Chaplygin inflationary
model in the context of a branewold scenario. In the slow-roll
approximation we have found a general relation between the scalar
potential and its derivative. This has led us to a general
criterium for inflation to occur (see Eq.(\ref{cond})). We  have
also obtained explicit expressions for the corresponding scalar
spectrum index $n_s$ and its running $\alpha_s$.

 By using  an exponential  potential in the high-energy
regime  and from the WMAP five year data,  we found the
constraints  of the parameters $A$ and $\alpha$ from the
tensor-scalar ratio $r$ (see Figs. 2 and 3).  In order to bring
some explicit results we have taken the constraints $A\sim
10^{-25} m_p^8$ and $\alpha\sim 10^{-6}m_p$, from which we get the
values $V_*\sim 10^{-12}m_p^4$, $V_f\sim 10^{-13}m_p^4$,
$\lambda\sim 10^{-13}m_p^4$ and $N\sim53$. Here, we have used  the
WMAP five year data where $P_{\cal R}(k_*)\simeq 2.4\times
10^{-9}$, $n_s(k_*)\simeq 0.96$, $\alpha_s(k_*)\simeq-0.03$ and
$r(k_*)\simeq 0.055$. Note that the restrictions imposed by
currents observational data allowed us to establish a small range
for the parameters that appear in the tachyon-brane-Chaplygin
inflationary model.

We have not addressed reheating and transition to standard
cosmology in our model (see e.g., Ref.\cite{u}). Specifically, it
will be very interesting to know how
 the reheating temperature
in the hight-energy scenario, contributes to establish   some
constrains on the parameters of the model. We hope to return to
this point in the near future.

\begin{acknowledgments}
  This work was supported by the ``Programa Bicentenario de
Ciencia y Tecnolog\'{\i}a" through the Grant \mbox {N$^0$ PSD/06}.
\end{acknowledgments}


\end{document}